\documentclass[journal=jacsat,manuscript=article]{achemso}

\SectionNumbersOn
\usepackage{tabularx}
\usepackage{amsfonts,amsmath,subcaption}
\usepackage{chemformula} 
\usepackage[T1]{fontenc} 

\author{Taehee Ko}
\affiliation{Department of Mathematics, Penn State University, University Park, PA 16802, USA}
\email{tuk351@psu.edu}
\author{Joseph Heindel}
\altaffiliation{Chemical Sciences Division, Lawrence Berkeley National Laboratory, Berkeley, CA 94720, USA}
\email{heindelj@lbl.gov}
\author{Xingyi Guan}
\altaffiliation{Chemical Sciences Division, Lawrence Berkeley National Laboratory, Berkeley, CA 94720, USA}
\email{nancy_guan@berkeley.edu}
\author{Teresa Head-Gordon}
\affiliation{Kenneth S. Pitzer Theory Center and Department of Chemistry, University of California, Berkeley, CA 94720, USA}
\altaffiliation{Chemical Sciences Division, Lawrence Berkeley National Laboratory, Berkeley, CA 94720, USA}
\alsoaffiliation{Departments of Bioengineering and Chemical and Biomolecular Engineering, University of California, Berkeley, CA 94720, USA}
\email{thg@berkeley.edu}
\author{David Williams-Young}
\author{Chao Yang}
\affiliation{Applied Mathematics and Computational Research Division, Lawrence Berkeley National Laboratory, Berkeley, CA 94720, USA}
\email{dbwy@lbl.gov, CYang@lbl.gov} 

\title[An \textsf{achemso} demo]
  {Using Diffusion Maps to Analyze Reaction Dynamics for a Hydrogen Combustion Benchmark Dataset}

\abbreviations{IR,NMR,UV}

\begin{document}

\begin{tocentry}

Some journals require a graphical entry for the Table of Contents.
This should be laid out ``print ready'' so that the sizing of the
text is correct.

Inside the \texttt{tocentry} environment, the font used is Helvetica
8\,pt, as required by \emph{Journal of the American Chemical
Society}.

The surrounding frame is 9\,cm by 3.5\,cm, which is the maximum
permitted for  \emph{Journal of the American Chemical Society}
graphical table of content entries. The box will not resize if the
content is too big: instead it will overflow the edge of the box.

This box and the associated title will always be printed on a
separate page at the end of the document.

\end{tocentry}



\begin{abstract}
We use local diffusion maps to assess the quality of two types of collective variables (CVs) for a recently published hydrogen combustion benchmark dataset~\cite{guan2022benchmark} that contains ab initio molecular dynamics trajectories and normal modes along minimum energy paths. This approach was recently advocated in~\cite{tlldiffmap20} for assessing CVs and analyzing reactions modeled by classical molecular dynamics simulations. We report the effectiveness of this approach to molecular systems modeled by quantum ab initio molecular dynamics.  In addition to assessing the quality of CVs, we also use global diffusion maps to perform committor analysis as proposed in~\cite{tlldiffmap20}.  We show that the committor function obtained from the global diffusion map allows us to identify transition regions of interest in several hydrogen combustion reaction channels.
\end{abstract}

\section{Introduction}
The search for effective collective variables (CVs) for many-body molecular systems is paramount for characterizing the primary dynamic pathway(s) for conformational changes and chemical transformations in a tractable and/or more physically interpretable lower manifold.\cite{fiorin2013using, noe2017collective,karmakar2021collective,mendels2018collective} For some molecular systems such as the alanine dipeptide, physically intuitive CVs such as dihedral angles perform very well, and for chemical reactions simple CVs based on distances are often assumed to be adequate. In such cases they can also be readily combined with enhanced sampling techniques such as metadynamics\cite{barducci2011metadynamics,sutto2012new} or umbrella sampling\cite{kastner2011umbrella,you2019potential}.

Several methods have been developed to identify CVs, including those using machine learning\cite{belkacemi2021chasing}, but they are often validated by using molecular systems for which a lot is known already, especially for the alanine dipeptide\cite{bonati2020data,mori2020learning,sultan2018automated,kikutsuji2022explaining} However, in general it is not so clear how to identify relevant CVs for complex chemical or materials systems in the condensed phase or when chemical reactions occur. For example, aqueous phase transformations can involve many solvent molecules that undergo concerted rearrangements that are mediated through long-range interactions.\cite{ma2005automatic}  Furthermore, these studies usually utilize classical molecular dynamics (MD) which typically limits one to studying only conformational changes. For chemical reactions involving bond forming and breaking, a quantum mechanical treatment is needed but can be computationally slow or prohibitive and the CVs can be unintuitive for complex reaction mechanisms. 

Recently, diffusion maps have been shown to be an effective tool for identifying good collective variables and computing committor functions that can be used to elucidate chemical reactions in molecular systems~\cite{tlldiffmap20}. The diffusion map or its variants have been widely used in many MD  applications\cite{tlldiffmap20,rohrdanz2011determination,boninsegna2015investigating,coifman2008diffusion} and relevant to the topic explored here, for searching for rare transitions within dynamical information.\cite{coifman2008diffusion,tlldiffmap20,ferguson2011integrating} While the diffusion coordinates obtained from the eigenvectors of the diffusion map operator have been proposed as a means to obtain CVs, their interpretation in terms of the potential energy surface (PES) can be unclear. We will use the technique recently proposed in~\cite{tlldiffmap20} to evaluate and compare CVs obtained from two commonly used approaches, i.e, internal coordinates and principal component analysis. We will also examine the possibility of using diffusion maps to compute committor functions which describe the probability of reaching either of two local minima from a particular configuration using an ensemble of short trajectories\cite{bolhuis2002transition,khoo2019solving}. The committor function is of central importance because it generalizes the concept of a transition state by explicitly accounting for dynamics on a high-dimensional PES. 

We apply the diffusion map technique to a hydrogen combustion benchmark dataset that contains ab initio molecular dynamics (AIMD) trajectories and normal modes along minimum energy paths\cite{guan2022benchmark}, and report the effectiveness of this approach to reaction dynamics. This dataset is ideal for analyzing diffusion maps and commitor functions as the molecular species are relatively small, the energies (and forces) are generated from reliable density functional theory (DFT) using the $\omega$B97X-V DFT functional\cite{Mardirossian2014} with the cc-pVTZ basis set, and configurations near reaction barriers as well as configurations in metastable regions are well-sampled for a variety of reaction channels. 

The paper is organized as follows. In section~\ref{sec:diffmap}, we review several techniques to construct a diffusion map. In section~\ref{sec:LDQSD}, we introduce the mathematical description of the overdamped Langevin dynamics and the quasi-stationary distribution (QSD) which is the required distribution of molecular configurations used to construct an effective diffusion map. In section~\ref{sec:committor}, we describe an algorithm to compute committor probabilities based on the diffusion map. We review two commonly used approaches for constructing CVs in section~\ref{sec:dmcvs} and provide details on the hydrogen combustion dataset in section~\ref{sec:HComb}. In section~\ref{sec:results} of results, we report the good CVs for all hydrogen combustion reactions identified by diffusion maps, and validate such an assessment by examining the PES of four representative reactions in different pairs of CVs. We also show the committor functions obtained from global diffusion maps for four representative reactions and discuss how well they characterize the reaction mechanism of these reactions. Finally we offer concluding remarks and future directions in Section 4.

\section{Theory and Methods}


\noindent
\subsection{Diffusion coordinates and diffusion map construction}\label{sec:diffmap}

We consider an overdamped Langevin dynamics
\begin{equation}\label{Langevin}
    dx_t = -\nabla U(x_t)dt+\sqrt{2\beta^{-1}}dW_t,
\end{equation}
where $x$ is a molecular configuration, $U(x)$ is the potential energy at $x$ that takes into account quantum mechnical forces among different atoms, $\beta$ is the inverse temperature and $W_t$ is a standard Wiener process. The infinitesimal generator of the Markov process $(x_t)_{t\geq 0}$ associated with \eqref{Langevin} is related to the Kolmogorov operator
\begin{equation}\label{Loperator}
    L = -\nabla U\cdot \nabla+\beta^{-1}\Delta .
\end{equation}For example, for any smooth observable $f:\Omega\rightarrow\mathbb{R}$ in $L_2(\Omega)$, the expected value of $f(x_t)$ along a trajectory governed by the overdamped Lagenvin dynamics \eqref{Langevin}, denoted by $g(x,t):=\mathbb{E}[f(x_t)]$, satisfies the backward Kolmogorov equation
\begin{equation}
    \frac{\partial g}{\partial t}= Lg,
    \label{eq:fpeq}
\end{equation}
where the region $\Omega$ is compact, the elliptic operator $L$ has a discrete set of eigenvalues $\lambda_j$ and the corresponding eigenfunctions $\psi_j(x)$.  The solution to \eqref{eq:fpeq} can be expressed in terms of eigenpairs $(\lambda_j,\psi_j(x))$ of $L$, i.e., \begin{equation} g(x,t)  = \sum_{j=1}^{\infty}b_je^{-\lambda_jt}\psi_j(x),
\end{equation}
where the coefficient $b_j$ is derived from the duality with the forward Kolmogorov operator \cite{coifman2008diffusion} and the eigenvalues of $L$ satisfy that $0=\lambda_1<\lambda_2\leq\lambda_3\leq...$. The nonzero eigenvalues and the corresponding eigenfunctions of $L$ evaluated at $x$ can be used to define a set of \emph{diffusion coordinates} (DC)
\begin{equation}\label{eq:contDCs}
    \left(e^{-\lambda_2t}\psi_2(x),e^{-\lambda_3t}\psi_3(x),...\right).
\end{equation}
As a result, the distance between $x$ and $y$ can be measured in terms of the $L_2$ norm of the diffusion coordinates associated with $x$ and $y$. This is often referred to as the \emph{diffusion distance} $D_t^{(x,y)}$,
where
\begin{equation}
    D_t^2(x,y) = \sum_{j\geq 2}e^{-2\lambda_jt}|\psi_j(x)-\psi_j(y)|^2.
    \label{eq:diffdist}
\end{equation}

\noindent
If there is a large spectral gap between the $\lambda_{k+1}$ and $\lambda_{k+2}$, the diffusion coordinate \eqref{eq:contDCs} is dominated by the first $k$ components
\begin{equation}\label{eq:truncDCs}
    \left(e^{-\lambda_2t}\psi_2(x),e^{-\lambda_3t}\psi_3(x),...,e^{-\lambda_{k}t}\psi_{k}(x),e^{-\lambda_{k+1}t}\psi_{k+1}(x)\right).
\end{equation}
As a result, the diffusion distance $D_t^{(x,y)}$ can be computed from \eqref{eq:diffdist} by keeping the first $k$ terms.
The use of a $k$-component diffusion coordinate for a relatively small $k$ allows us to achieve significant dimension reduction.

However, finding the diffusion coordinates in high dimensions by computing the eigenvalues and eigenvectors of a discretized (by, e.g., a finite-element method) $L$ defined in \eqref{Loperator} is computationally intractable in general.
Alternatively, one can obtain approximate diffusion coordinates \eqref{eq:contDCs} by constructing a transition probability matrix from configurations sampled along a trajectory using a Gaussian kernel. This kernel is referred to as a diffusion map \cite{tlldiffmap20,coifman2005geometric}. In the following, we briefly describe the basic steps of a diffusion map construction and how diffusion coordinates can be obtained from a diffusion map.

The initial step in the construction of a diffusion map associated with the operator \eqref{Loperator} is to build a kernel matrix from configurations $\{x_i\}_{i=1}^n$ sampled along a trajectory of an overdamped Lagevin dynamics~\eqref{Langevin}, which is of the form
\begin{equation}\label{eq:conste}
    [A_{\epsilon}]_{ij} = \exp\biggl(-\frac{\|x_i-x_j\|^2}{\varepsilon}\biggr).
\end{equation}
The local parameter $\varepsilon$ scales as distances between the samples. The kernel matrix $A_{\epsilon}$ can be further normalized by, for example, configuration-dependent densities or the row sums of the kernel matrix, to yield \cite{coifman2008diffusion}
\begin{equation}\label{eq:kernelmat1}
[\hat{A}_{\epsilon,\alpha}]_{ij}:=\frac{[A_{\epsilon}]_{ij}}{p_i^{\alpha}p_j^{\alpha}},
\end{equation}
where $p_i:=\sum_{k=1}^n[A_{\epsilon}]_{ik}$ is the $i$th row sum of $A$ and $\alpha \in [0,1]$ is an appropriate chosen parameter. 

To obtain a transition probability matrix $P_{\varepsilon,\alpha}$ from $\hat{A}_{\varepsilon,\alpha}$, we scale it by a diagonal matrix $D$, i.e.,
\begin{equation}
\label{eq:Pmatrix}
P_{\varepsilon,\alpha} = D^{-1}\hat{A},
\end{equation}
where the $i$th diagonal entry of $D$ equals the $i$th row sum of $\hat{A}_{\varepsilon,\alpha}$. Importantly, 
in the limit of $n\rightarrow \infty$ and $\varepsilon\rightarrow 0$,
\begin{equation}\label{eq:asympdiff}
    \frac{P_{\epsilon,\alpha}-I}{\epsilon}\rightarrow L.
\end{equation}
This expression allows us to construct an approximation to $L$ from the transition probability matrix of sampled configurations along a Langevin trajectory.\cite[Theorem 2]{coifman2006diffusion} Therefore, we can obtain approximate diffusion coordinates \eqref{eq:truncDCs} by computing eigenvalues and eigenvectors of  the matrix on the left-hand side of \eqref{eq:asympdiff}. The parameter $\alpha\in[0,1]$ determines the type of the continuous operator on the right hand side of \eqref{eq:asympdiff} in the limit $n\rightarrow\infty$ and $\epsilon\rightarrow 0$. In particular, when $\alpha = \frac{1}{2}$, $\hat{A}_{\epsilon,\alpha}$ converges to the backward Kolmogorov operator that appears in \eqref{Loperator} and \eqref{eq:fpeq} (see \cite{coifman2005geometric}). 


Several variations of the diffusion map have been developed to improve the effectiveness of dimension reduction through the use of diffusion coordinates. For molecular dynamics simulations, one can use the energy-based kernel suggested in \cite{coifman2008diffusion},  
\begin{equation}\label{eq:kernelmat2} [\hat{A}]_{ij}:=\frac{[A_{\varepsilon}]_{ij}}{\sqrt{e^{-\beta U(x_i)}e^{-\beta U(x_j)}}},
\end{equation}
where $U(x)$ represents the energy of the configuration $x$ and $\beta$ is the inverse temperature. A generalization of this kernel was established and proven theoretically to ensure a small set of diffusion coordinates can capture the underlying manifold $\mathcal{M}$  in \cite{berry2016variable}. Nevertheless, in situations where sampling is highly varied due to some structural instability, finding a proper scaling parameter $\varepsilon$ in the kernel \eqref{eq:conste} can be difficult due to large variations in local scales associated with the sampling. In such situations, it may be better to define a kernel matrix with configuration-dependent scaling parameters $\{\varepsilon_i\}_{i=1}^n$ instead of a single constant  \cite{rohrdanz2011determination,preto2014fast}, for example,
\begin{equation}\label{eq:varye}
    [A_{\{\varepsilon\}_{i=1}^n}]_{ij} = \exp\biggl(-\frac{\|x_i-x_j\|^2}{\varepsilon_i\varepsilon_j}\biggr).
\end{equation}
A simple rule of selecting those configuration-dependent scaling parameters is to use the nearest neighborhood criterion, i.e.,
\begin{equation}\label{knearest}
    \varepsilon_i = \textrm{the }rn\textrm{-th smallest value in the distances }\{\|x_i-x_j\|\}_{j=1}^n,
\end{equation}where the parameter $0<r<1$ determines the size of neighborhood. A range of procedures for selecting the local scales have been proposed \cite{preto2014fast,little2009multiscale}. 

Finally, one can apply the normalizations \eqref{eq:kernelmat1} and \eqref{eq:kernelmat2} to the locally-scaled kernel \eqref{eq:varye}, which yields more robust diffusion maps than the counterparts with the constant scale kernel \eqref{eq:conste}. This is formulated as the \emph{weighted} kernel matrix,
\begin{equation}\label{eq:kernelmat3}
    [\hat{A}_{\{\varepsilon\}_{i=1}^n}]_{ij} = \sqrt{w_iw_j}\exp\biggl(-\frac{\|x_i-x_j\|^2}{\varepsilon_i\varepsilon_j}\biggr).
\end{equation}As demonstrated in \cite{preto2014fast}, 
the weights $w_i$ play a role to correct a bias from simulation and help recover the \emph{unbiased} dynamical information such as free energy. 

\subsection{Local diffusion map and quasi-stationary distribution}\label{sec:LDQSD}
Trajectories from the Langevin dynamics \eqref{Langevin} tend to stay in a metastable region for a very long time before exiting the region. It has been suggested that diffusion coordinates obtained from a ``local" diffusion map constructed from samples along such a trajectory can be used to identify high-quality CVs as long as these samples satisfy a so-called quasi-stationary distribution (QSD) \cite{tlldiffmap20,le2012mathematical}.  

By definition, a QSD for the stochastic process \eqref{Langevin} is a probability measure $\nu$ on a metastable region $\Omega$ that satisfies 
\begin{equation}\label{QSD1stp}
    \lim_{t\rightarrow \infty}\mathbb{E}[f(x_t)|T>t]=\int_{\Omega}f(x)\nu(dx),
\end{equation}
for any initial configuration $x_0\in\Omega$ and any smooth function $f:\mathbb{R}^d\rightarrow\mathbb{R}$ and the random variable $T$ is the first exit time of the process $x_t$, i.e.,
\begin{equation}
    T:=\inf\{t\geq 0 : x_t\not\in\Omega\}.
\end{equation}
It follows from \eqref{QSD1stp} that the expected value of $f(x)$ in $\Omega$ with respect to the measure $\nu$ can be approximately obtained from the expected value of $f(x_t)$ along a sufficiently long Langevin trajectory that stays within $\Omega$.


A practical question we need to address in using samples along a Langevin trajectory to construct a local diffusion map is to ensure that these samples are within a metastable region $\Omega$ and satisfy a QSD. A practical procedure for achieving such a goal is based on the connection between the first eigenfunction $u_1$ of the Kolmogorov operator (Eq. \eqref{Loperator}) and $\nu(x)$
described by the equation \cite{le2012mathematical},  
\begin{equation}\label{nu}
    \nu(x) = \frac{u_1(x)e^{-\beta V(x)}}{\int_{\Omega}u_1(x)e^{-\beta V(x)}dx} \textrm{ for all }x\in\Omega.
\end{equation}

The expression given in \eqref{nu} suggests that we can determine whether samples along a Langevin dynamics satisfy a QSD by monitoring the first eigenvalue of the operator $L$ associated with the diffusion map constructed from these samples.  When the eigenvalue does not change much, we can consider the samples to satisfy a QSD. In practice, we construct diffusion maps from a trajectory data iteratively and keep track of their first few eigenvalues instead of just the first one to determine whether samples along a trajectory satisfy the QSD.

\subsection{Global diffusion map and committor function}\label{sec:committor}The diffusion map is an effective tool for computing the committor function \cite{tlldiffmap20,prinz2011efficient}.  For two metastable regions $A$ and $B$, the committor function assigns to snapshot $x\in\Omega$ the probability of a trajectory starting from $x$ to reach $B$ first rather than $A$, namely,
\begin{equation}\label{committor}
    q(x):=\mathbb{P}\{\tau_B<\tau_A|x_0=x\},
\end{equation}where the random variable $\tau_A$ ($\tau_B)$ is the first time when a trajectory initialized at $x_0$ hits $A$ ($B)$. Note that $q(x)=0$ if $x\in A$ and $1$ if $x\in B$. Ideally, as a trajectory gets closer to $B$ from $A$, the committor function continuously increases from $0$ to $1$ and the isocommittor surfaces ($q(x)=0.5$) can be used to define transition regions \cite{weinan2005transition}.  

On a global region $\Omega$ that includes $A$ and $B$, the committor function can be 
  interpreted as a solution to the backward Kolmogorov equation with a boundary condition, 
\begin{equation}
\label{eq:Leq}
    \begin{cases}
    Lq = 0, x\in \Omega-(A\cup B)\\
    q(x) = 0,\; x\in A\\
    q(x) = 1,\; x\in B
    \end{cases},
\end{equation}where $L$ is the operator defined in \eqref{Loperator}. To find an approximate solution, one can exploit the asymptotic property of the diffusion map \eqref{eq:asympdiff} and can compute an approximate committor function $q$ by solving the following system of linear equations \cite{prinz2011efficient,tlldiffmap20},
\begin{equation}\label{eq:committor}
    (P_{\varepsilon,\alpha}-I)[c,c]q[c] = -(P_{\varepsilon,\alpha}-I)[c,b]q[b],
\end{equation}
where $c$ and $b$ represent the set of indices for configurations in the complement of $A \cup B$ and in the metastable region $B$, respectively. Furthermore, if local distance scales are highly variable for given snapshots, one can use the kernel defined with different local scales \eqref{eq:kernelmat3} for solving the equations \eqref{eq:committor}.

\subsection{Dimension reduction and the choice of CVs}
\label{sec:dmcvs}
There are many ways to choose CVs. In this work, we focus on two particular types of CVs, internal coordinates and principal component analysis, that are commonly used to define reaction coordinates of molecular systems. 
Internal coordinates (ICs) refer to bond lengths, bond angles, and dihedral angles of a bonded molecule. Compared to Cartesian coordinates, ICs have the desirable advantage that they are invariant under an overall translation or rotation of the molecule which do not alter the potential energy of the molecule. Selected ICs are sometimes used as CVs. A well-known example is the use of two dihedral angles of an alanine dipetide molecule as CVs to examine conformational changes of the molecule \cite{tlldiffmap20,preto2014fast}. These will be examined with respect to diffusion coordinates in Section 7.

Principal component analysis (PCA) is one of the most common and useful dimensionality reduction techniques to represent complex information as low dimensional data. We apply the PCA to identify directions in which the data are varied most, (e.g. the projection of data onto the 1st principal component (PC) with the largest variance, the second largest variance in the 2nd PC, and so on). Given $X\in\mathbb{R}^{3n\times N}$, where $N$ is the number of samples, the principal components (PCs) are defined as the eigenvectors of the covariance of $X$ or the right-singular vectors via the singular value decomposition (SVD),
\begin{equation}\label{PCA}
    X-\bar{X}=Y\Sigma Z^T,
\end{equation}
where $\bar{X}$ is the mean over all samples, column vectors of $Y$ are the PCs and the diagonal matrix $\Sigma$ contains the singular values of $X-\bar{X}$. Each configuration can be expanded as a linear combination of the PCs contained in $Y$. The coefficients of the $j$th PC ($y_j$) for all configurations contained in $X$ can be obtained from
\begin{equation}\label{eq:pcaC}
C_j = X^T y_j.    
\end{equation}

If the first few singular values in $\Sigma$ are much larger than other singular values, each configuration in $X$ can be well represented by the first few PCs.  The coefficients associated with these PCs can be used as CVs. 

\subsection{Hydrogen combustion dataset}\label{sec:HComb}
We consider the effectiveness of using a local diffusion map to assess the quality of CVs obtained for several reactions in the hydrogen combustion benchmark data published in~\cite{hcombustdata22}. Table~\ref{tb:reactions} lists all 19 reactions contained in the hydrogen combustion benchmark data.\cite{hcombustdata22} The benchmark data contains molecular configurations sampled along a MEP for each reaction, including the reactant, transition, and product states. The dataset contains 290,000 potential energies and 1,270,000 forces for hydrogen compounds in different molecular configurations. These configurations are generated through normal mode sampling and ab initio molecular dynamics (AIMD) at different temperatures beginning from different points along the 0K intrinsic reaction coordinate (IRC) or equivalently MEP, including the transition state. In regards the AIMD simulations for the 19 reaction channels, each reaction has 10000 snapshots obtained at four different temperatures: 500K, 1000K, 2000K, and 3000K generated from running AIMDs (using the Q-Chem software pacakage\cite{Epifanovsky2021} with the transition state as the starting point. As reported in~\cite{hcombustdata22}, some of the reactions are relatively simple, e.g., reactions 5 and 6, involving the association
and dissociation of two atoms along certain directions which constitutes the only degree of freedom; we will not examine this type of reaction in this paper. Among the 19 reaction channels, we focus on reactions 09, 11, 14, and 16 as representatives for association, O-transfer,  H-transfer, and substitution, respectively. These reactions also include molecules with more atoms such that the number of degrees of freedom in ICs can be as large as 18.

\begin{table}[htbp]
\begin{center}
\begin{tabular}{ | m{6cm} | m{1cm}| m{1cm} | m{1cm} |}
  \hline 
  No. Reaction & Atoms & $\mathrm{DoF}$ & $\mathrm{DoF_{int}}$ \\ 
  \hline
  \textbf{Association/Dissociation} & & & \\
  5. H$_2\longrightarrow \;$2H & 2 & 6 & 1 \\
  6. O$_2\longrightarrow \;$2O & 2 & 6 & 1 \\
  7. OH$\;\longrightarrow \;$O+H & 2 & 6 & 1\\
  8. H+OH $\longrightarrow$ H$_2$O & 3 & 9 &  3 \\
  9. H+O$_2 \longrightarrow \;$HO$_2$ & 3 & 9 & 3 \\
  15. H$_2$O$_2$ $\longrightarrow\;$2OH & 4 & 12 & 6 \\  
  \hline
  \textbf{Substitution} &&&\\
  16. H$_2$O$_2$+H $\longrightarrow\;$H$_2$O+OH & 5 & 15 & 9 \\
  \hline
  \textbf{O-transfer} &&&\\
  1. H+O$_2 \longrightarrow\; $OH+O & 3 & 9 & 3 \\
  11. HO$_2$+H$\;\longrightarrow\;$2OH & 4 & 12 &  6 \\
  12. HO$_2$+O$\;\longrightarrow\;$OH+O$_2$ & 4 & 12 &  6 \\ 
  \hline
  \textbf{H-transfer} &&&\\
  2. O+H$_2 \longrightarrow\;$OH+H& 3 & 9 & 3 \\
  3. H$_2$+OH $\longrightarrow\;$H$_2$O+H & 4 & 12 & 6\\
  4. H$_2$O $\longrightarrow\;$2OH & 4 & 12 & 6 \\
  10. HO$_2$+H $\longrightarrow\;$H$_2$+O$_2$ & 4 & 12 & 6 \\
  13. HO$_2$+OH $\longrightarrow\;$H$_2$O+O$_2$& 5 & 12 & 9 \\
  14. 2HO$_2\longrightarrow\;$ H$_2$O$_2$+O$_2$& 6 & 18 & 12 \\
  17. H$_2$O$_2$+H $\longrightarrow\;$HO$_2$+H$_2$& 5 & 15 & 9 \\
  18. H$_2$O$_2$+O $\longrightarrow\;$ HO$_2$+OH & 5 & 15 & 9 \\
  19. H$_2$O$_2$+OH $\longrightarrow\;$H$_2$O+HO$_2$& 6 & 18 & 12 \\
  \hline
\end{tabular}
\end{center}
\caption{The 19 reactions contained in the hydrogen combustion benchmark dataset. The number of atoms involved in each reaction, the total number of degrees of freedom ($\mathrm{DoF}$) in Cartesian coordinates, and total number of degrees of freedom in ICs ( $\mathrm{DoF}_{int}$.)}
\label{tb:reactions}
\end{table}

\noindent

We should also note that in~\cite{hcombustdata22}, the minimum energy path (MEP) is plotted in terms of two coordination numbers (CNs) that are considered as internal reaction coordinates (IRC). A coordination number is defined as the number of bonded or closest neighbors of a central atom or molecule of interest. In the hydrogen combustion study\cite{hcombustdata22}, the $i$th atom in a molecule is defined as
\begin{equation}
CN_i = \sum_{j\neq i} \frac{2.0}{1+\exp(\sigma(r_j-r_{0,j}))},
\label{eq:cn}
\end{equation}
where $r_j$ is the distance between atom $i$ and atom $j$ and $r_{0,j}$ is the equilibrium distance between atom $i$ and $j$. Each Fermi-Dirac function in \eqref{eq:cn} is close to zero if $r_j - r_{0,j}$ is large, and close to 1 if $r_j - r_{0,j}$ is small. Therefore, \eqref{eq:cn} effectively gives the number of neighboring atoms of atom $i$ that are close to be in equilibrium positions. In~\cite{hcombustdata22}, one or two coordination numbers were chosen as CVs that incorporated the IRC MEP that connects a local minimum of the potential energy surface (reactant) to another local minimum (product).

To generate additional samples near the reactant and product states (local minima), which are often available in a practical setting, we ran some additional MD simulations using QChem\cite{Epifanovsky2021} starting from either the reactant or product state. We used density functional theory (DFT), specifically the $\omega$B97X-V functional and the cc-pVTZ basis set to perform potential energy calculations and molecular dynamics simulations. For all molecular dynamics simulations, we performed Langevin dynamics at 300K with a 0.12 femtosecond (fs) time step; we set the reactant provided in the IRC data \cite{hcombustdata22} as the initial configuration of the dynamics. For reactions 11, 14, and 16, we generated 4000 snapshots. We could only generate 105 snapshots for reaction 09 before dissociation occurred.

\section{Results}\label{sec:results}

\subsection{Checking QSD for local diffusion map construction}
As we discussed in section~\ref{sec:diffmap}, a local diffusion map constructed from molecular configurations within a metastable region can be used to assess the quality of CVs as long as the configurations used to construct the diffusion map satisfies a QSD \cite{collet2013quasi}. 

\begin{figure}[htbp]
   \includegraphics[width=0.99\textwidth]{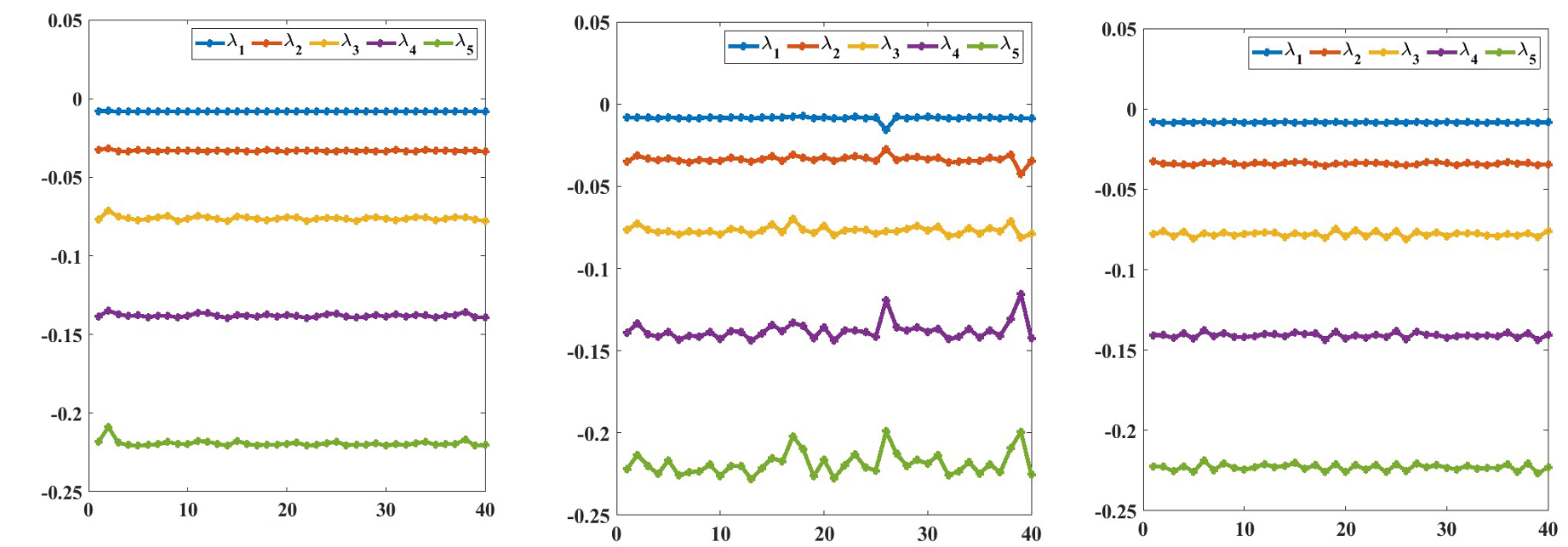}
\center
\caption{The changes of the first few dominant eigenvalues of the diffusion maps associated with reactions 11 (left), 14 (middle) and 16 (right). The diffusion maps are constructed every $m=100$ AIMD steps of a 4000-step trajectory from snapshots sampled along the trajectory.}
    \label{fig:qsd11}
\end{figure}

Due to the connection between the distribution of configurations along an MD trajectory and the eigenfunction associated with the first eigenvalue of corresponding Kolmogorov operator \eqref{Loperator} as discussed in section~\ref{sec:diffmap}, we can check whether the sampled MD snapshots satisfy QSD by monitoring the spectrum of the diffusion map constructed from these snapshots. To be specific, every $m=100$ MD steps, which constitutes an iteration of the QSD checking procedure, we construct a diffusion map using samples along the trajectories available up that point and compute the first few eigenvalues of the corresponding approximate Kolmogorav operator $L_{\varepsilon}$. These eigenvalues are shown in Figure~\ref{fig:qsd11} for reactions 11, 14 and 16. Because each MD trajectory consists of $N=4000$ snapshots, $N/m=4000/100=40$ iterations were performed to monitor the change in the eigenvalues of $L_{\varepsilon}$.

Figure~\ref{fig:qsd11} shows that the first 5 eigenvalues of the diffusion map for reaction channels 11 and 16 do not change much during the entire simulation. For reaction channel 14, the dominant eigenvalues of the diffusion map begin to fluctuate around a few mean values, indicating that the subsequent trajectory samples no longer fulfill the QSD defined on the metastable region. Based on those measures, we pick the subset of the snapshots up to the point when the first 5 eigenvalues of the diffusion map begin to change more significantly. The configurations within such a subset are deemed to statisfy QSD. For reaction 9, because there are only 105 snapshots in the AIMD trajectory, we use all of them because they appear to be within a metastable region.

 
\subsection{Implementation details}
In all experiments, we used ICs of sampled configurations to construct local diffusion maps. For global diffusion maps used in section~\ref{sec:commanalysis} to perform committor analyses, we found that root mean squared deviation (RMSD) aligned Cartesian coordinates, as defined in \cite{kearsley1989orthogonal}, were sometimes more effective. 
When using ICs to construct a diffusion map kernel matrix, it is important to note that the absolute difference in two angles should never be larger than $\pi$ when the difference of two ICs are used to evaluate a kernel matrix element defined in \eqref{eq:conste}.  For example, if $\theta_1=\pi/6$ and $\theta_2 = 2\pi -\pi/6$, the absolute difference between $\theta_1$ and $\theta_2$ should be $\pi/3$ instead of $2\pi-\pi/3$, i.e., when the difference between two angles exceeds $\pi$, we need to subtract $2\pi$ from the difference. Otherwise, the Euclidean difference between two sets of ICs can be artificially increased by the extra $\pi$ in the angle difference. This increase can yield distorted diffusion coordinates.

\begin{figure}[htbp]
\centering
\includegraphics[width=0.5\textwidth]{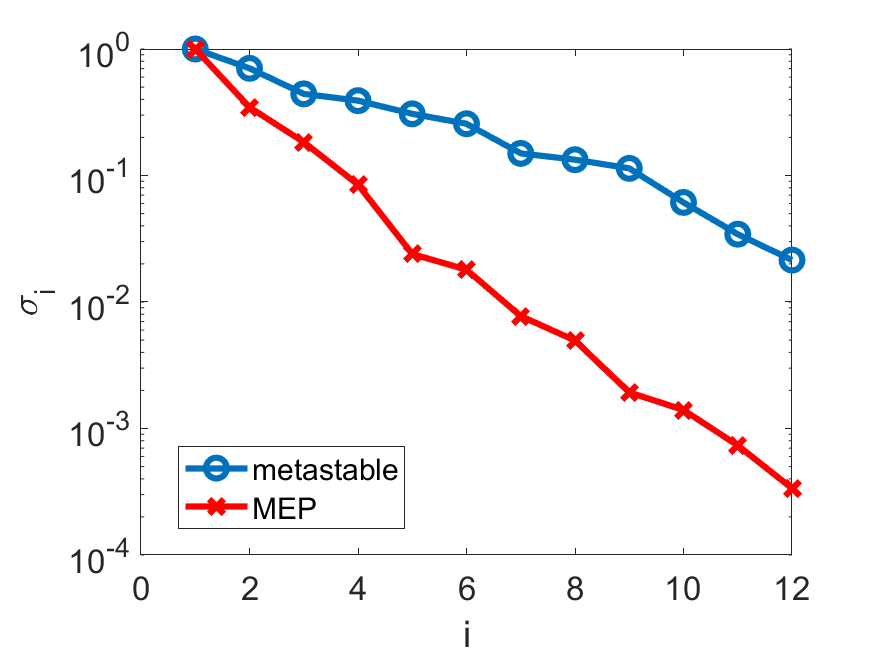}
    \caption{Normalized singular values of the matrix of configurations sampled within a metastable region (blue) and along the MEP (red) of reaction 19. The normalization is performed by dividing all singular values by the largest singular value.}
    \label{fig:svalues19}
\end{figure}

We perform PCA on sampled ICs. Ideally, we would like to use configurations along or close to the MEP because singular values resulting from PCA are likely to decrease faster and the dominant PCs obtained from such an analysis are likely to represent the main reaction mechanism well. 
For example, Figure~\ref{fig:svalues19} shows that the singular values obtained from the PCA performed on the configurations sampled along the MEP decreases much faster than those obtained from the configurations sampled within a metastable state of reaction 19.  However, because the MEP is unknown in general, this approach is not practical. In the following, we use MD snapshots sampled within a metastable region to perform the PCA even though such an analysis is not optimal in the sense that more principal components may be required to capture the reaction mechanism. 

In addition to using diffusion maps to access the quality of CVs, we validate the quality assessment by plotting the PES in these CVs to see if a saddle point can be observed on such a surface.  The presence of a saddle point would indicate that the chosen CVs are good CVs for describing the reaction mechanism near the transition state. To plot the PES with respect to two CVs $c_1$ and $c_2$, we evaluate the potential energy at
\begin{equation}
    z = z_{\mathrm{ts}}+c_1u_1+c_2u_2,
\end{equation}
where $z_{\mathrm{ts}}$ denotes coordinates of the transition state and $u_1,u_2$ are either two elementary basis vectors i.e., columns of a identity matrix when ICs are chosen as the CVs, or two PCs when PCA based CVs are to be examined. 
We plot the energies
\begin{equation}\label{eq:pes}
 E(c_1,c_2) \equiv E(z_{\mathrm{ts}}+c_1 u_1 + c_2 u_2).
\end{equation}
on a uniformed sampled 2D domain $[c_1^{\mathrm{lb}},c_1^{\mathrm{ub}}]\times [c_2^{\mathrm{lb}},c_2^{\mathrm{ub}}]$ for appropriately chosen
$c_1^{\mathrm{lb}}$, $c_1^{\mathrm{ub}}$, $c_2^{\mathrm{lb}}$ and $c_2^{\mathrm{ub}}$ values.

\subsection{Assessing the quality of CVs}
We construct the diffusion map and use it to   obtain two diffusion coordinates DC2 and DC3. We use these diffusion coordinates as a means to assess different definitions of CVs by examining the correlation of CVs with the diffusion coordinates. To be specific, we compute the Pearson correlation coefficient between CVs and the DCs as follows,   
\begin{equation}
\rho(C,D) = \frac{\mathrm{cov}(C,D)}{\sigma_C\sigma_D},
\label{eq:corrcoef}
\end{equation}
where $\mathrm{cov}(C,D)$ denotes the covariance between random variables $C$ and $D$, and $\sigma_C (\sigma_D)$ denotes the standard deviation of $C$ ($D$). If $C$ or $D$ are chosen as an IC, it can be easily obtained directly from the Cartesian coordinates of the snapshots used to construct the diffusion map. For the PCA, we use the variable in Eq. \eqref{eq:pcaC}.

Table \ref{tab:corr} reports the correlations of the IC and PCA based CVs with the first two diffusion coordinates (labelled as DC2 and DC3) for each representative reaction.  In the construction of the diffusion map, we set $r=0.1$ in \eqref{knearest}, and use the kernel \eqref{eq:kernelmat1} for (a) and (b) (3 and 4 atoms) and the energy-based kernel \eqref{eq:kernelmat2} for (c) and (d) (5 and 6 atoms). We observe that for each reaction several ICs can be highly correlated with the same DC. For example, for reaction 9, the absolute values of the  correlation coeffcient between IC2 and DC2, and between IC3 and DC2 are over 0.9, respectively. For reaction 11, IC3 and IC6 are highly correlated with DC2. For reaction 16, IC4 and IC9 are highly correlated with DC2. For reaction 14, IC2, IC3, IC6 and IC7 are all highly correlated with DC2. By contrast, we find that each DC tends to be highly correlated with only one of the PCA based CVs.


\begin{table}[htbp]
    \centering
    \begin{subtable}[h]{0.4\textwidth}
        \centering
        \begin{tabular}{l | l | l}
         & DC2 & DC3 \\
 \hline PC1 & {\color{red}0.94} & 0.29 \\
 \hline PC2 & 0.15 & 0.44 \\
 \hline PC3 & 0.29 & {\color{red}0.5} \\
 \hline IC1 (O1-O2) &  0.65 & 0.42 \\
 \hline IC2 (O1-H) & 0.94 & 0.29 \\
 \hline IC3 (H-O1-O2) & {\color{red}0.98} & 0.15 \\      \end{tabular}
       \caption{Association reaction 09}
    \end{subtable}
    \begin{subtable}[h]{0.4\textwidth}
        \centering
        \begin{tabular}{l | l | l}
         & DC2 & DC3 \\
       \hline PC1 & {\color{red}0.99} &  0.03 \\
 \hline PC2 & 0.08 & {\color{red}0.82} \\
 \hline PC3 & 0.12 & 0.44 \\
 \hline IC1 (H1-O1) & 0.00 & 0.01\\
 \hline IC2 (H1-O2) & 0.03 & 0.02\\
 \hline IC3 (H1-H2) & {\color{red}0.99} & 0.04 \\
 \hline IC4 (O2-H1-O1) & 0.06 & 0.01\\
 \hline IC5 (H2-H1-O1) & 0.62 & {\color{red}0.65}\\
 \hline IC6 (H2-H1-O1-O2) & 0.82 & 0.47 \\
        \end{tabular}
        \caption{O-transfer reaction 11}
     \end{subtable}
    
     \begin{subtable}[h]{0.4\textwidth}
        \centering
        \begin{tabular}{l | l | l}
         & DC2 & DC3 \\
       \hline PC1 & {\color{red}0.99} & 0.03 \\
 \hline PC2 & 0.08 & 0.28 \\
 \hline PC3 & 0.08 & {\color{red}0.93} \\
 \hline IC1 (H1-O1) & 0.01 & 0.00\\
 \hline IC2 (H1-O2) & 0.02 & 0.01\\
 \hline IC3 (H1-H2) & 0.01 & 0.04  \\
 \hline IC4 (O1-H3) & {\color{red}0.98} & 0.09\\
 \hline IC5 (O2-H1-O1) & 0.03 & 0.02\\
 \hline IC6 (H2-H1-O1) & 0.01 & 0.08\\
 \hline IC7 (H3-O1-O2) & 0.61 & {\color{red}0.77}\\
 \hline IC8 (H2-H1-O1-O2) & 0.01 & 0.08\\
 \hline IC9 (H3-O1-O2-H2) & 0.79 & 0.08\\
       \end{tabular}
       \caption{Substitution reaction 16}
    \end{subtable}
    \begin{subtable}[h]{0.4\textwidth}
        \centering
        \begin{tabular}{l | l | l}
         & DC2 & DC3 \\
        \hline PC1 & {\color{red}0.98}  & 0.12 \\
 \hline PC2 & 0.14 & {\color{red}0.94} \\
 \hline PC3 & 0.02 & 0.11 \\
 \hline IC1 (H1-O1) & 0.04 & 0.14 \\
 \hline IC2 (H1-O2) & 0.85 & 0.36 \\
 \hline IC3 (H1-H2) & 0.84 & 0.49 \\
 \hline IC4 (O1-O3) & 0.29 & {\color{red}0.72} \\
 \hline IC5 (O2-O4) & 0.04 & 0.23 \\
 \hline IC6 (O2-H1-O1) & 0.88 & 0.28 \\
 \hline IC7 (H2-H1-O1) & {\color{red}0.90} & 0.28 \\
 \hline IC8 (O3-O1-O2) & 0.1 & 0.69 \\
 \hline IC9 (O4-O2-H2) & 0.06 & 0.57 \\
 \hline IC10 (H2-H1-O1-O2) & 0.19 & 0.52 \\
 \hline IC11 (O3-O1-O2-H2) & 0.20 & 0.27 \\
 \hline IC12 (O4-O2-H2-O3) & 0.07 & 0.38 \\
        \end{tabular}
        \caption{H-transfer reaction 14}
     \end{subtable}
     \caption{The absolute values of correlation coefficients between ICs and PCA based CVs and DC2, DC3 for representative hydrogen combustion reaction channels 9, 11, 14 and 16. We highlight some of the relatively high correlation coefficients in red.}
     \label{tab:corr}
\end{table}

\begin{figure}[htbp]
   \centering
   \begin{subfigure}[b]{0.45\textwidth}
    \centering
    \includegraphics[width=\textwidth]{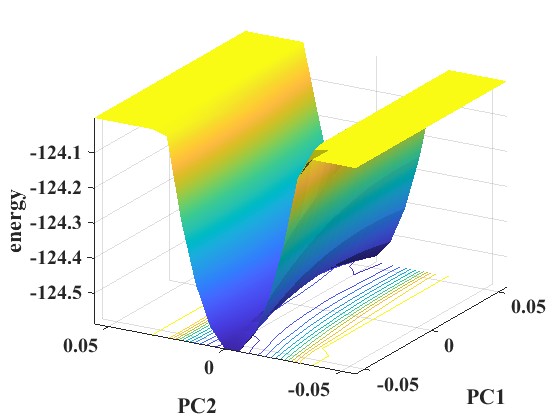}
    \caption{PC1, PC2}
    \end{subfigure}
   \begin{subfigure}[b]{0.45\textwidth}
    \centering
    \includegraphics[width=\textwidth]{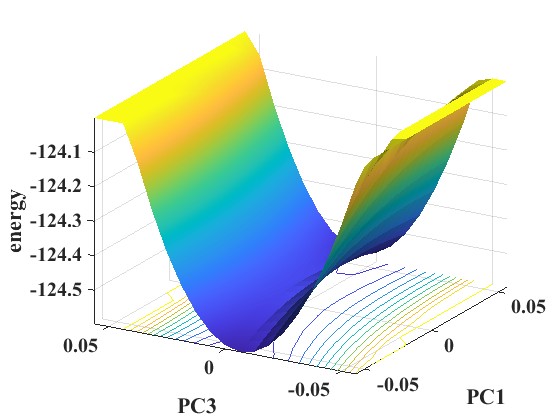}
    \caption{PC1, PC3}
    \end{subfigure}

    \begin{subfigure}[b]{0.45\textwidth}
    \centering
    \includegraphics[width=\textwidth]{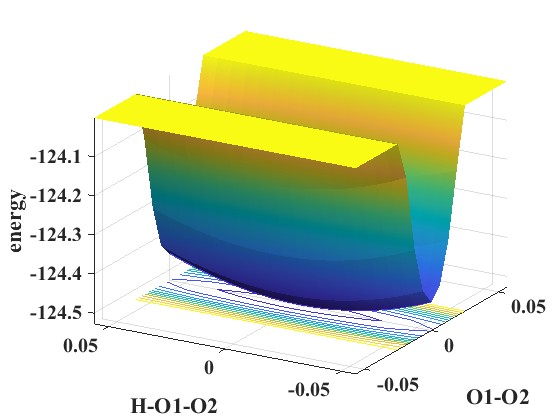}
    \caption{IC1(O1-O2), IC3(H-O1-O2)}
    \end{subfigure}
   \begin{subfigure}[b]{0.45\textwidth}
    \centering
    \includegraphics[width=\textwidth]{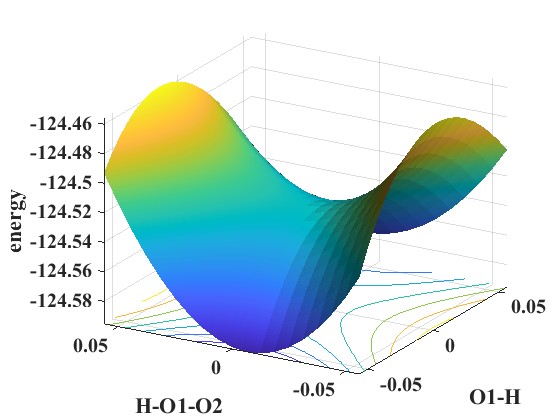}
    \caption{IC2(O1-H), IC3(H-O1-O2)}
    \end{subfigure}
    \caption{The potential energy surfaces of reaction 9 in pairs of PCs and pairs of ICs.}
    \label{fig:pes09}
\end{figure}

To validate the diffusion map based assessment of CVs shown in Table~\ref{tab:corr}, we plot potential energy surfaces in different combinations of CVs.  We evaluate the energy at each pair of CVs by using the formula \eqref{eq:pes} where $z_{\mathrm{st}}$ is the transition state specified in the hydrogen combustion dataset~ \cite{hcombustdata22}. For PCA based CVs, $u_1$ and $u_2$ are chosen to be the first and second principal component vectors or the first and the third ones.  For IC based CVs, $u_1$ and $u_2$ are standard basis vectors, i.e., columns of an identity matrix, with the position of 1 indicating which IC is chosen. When the projected transition state corresponds to a saddle point on this PES, we consider the CVs in which the PES is shown as good CVs.  
Figures \ref{fig:pes09} (a) and (b) show that that a saddle point can be found in the projected PES in both (PC1,PC2) and (PC1,PC3) for reaction 9.  PC1 appears to define the main reaction coordinate for reaction 9 along which the barrier is located at the full transition state.  This is consistent with the high correlation coefficient shown in Table~\ref{tab:corr}. PC2 and PC3 define the direction orthogonal to the reaction coordinate along which the barrier is a local minimum. We also observe from Figure~\ref{fig:pes09}(d) that the main reaction coordinate appears to be along IC2 which corresponds to the O1-H bond length. This is consistent with the relatively high correlation coefficient between IC2 and DC2 in Table~\ref{tab:corr} although the correlation coefficients between IC3 (H-O1-O2 angle) and DC2 is slightly higher.  A saddle point can clearly be seen in Figure~\ref{fig:pes09}(d) when the PES is plotted in terms of IC2 and IC3, whereas no such saddle point can be seen in Figure~\ref{fig:pes09}(c) where the PES is plotted in IC1 (O1-O2 bond length) and IC3 (H-O1-O2 angle). 

Furthermore, we observe that the 3 components of the first principal component, which correspond to the contribution of IC1, IC2 and IC3 to the principal component vector respectively are -0.088, 0.996, 0.0142. This indicates that IC2 contributes the most to the first principal component and consistent with the observation that the CVs defined in terms of PC1 and IC2 correspond to the same reaction coordinate. No clear saddle point can be seen in Figure~\ref{fig:pes09}(c) where the PES is plotted in terms of IC1 (the O-O bond distance) and IC3 (H-O1-O2 angle). This observation indicates that the O1-O2 bond distance is not a good CV for characterizing the reaction. This is also consistent with observation that IC1 has a relatively low correlation coefficient with DC2 and DC3.

\begin{figure}[htbp]
   \centering
   \begin{subfigure}[b]{0.45\textwidth}
    \centering
    \includegraphics[width=\textwidth]{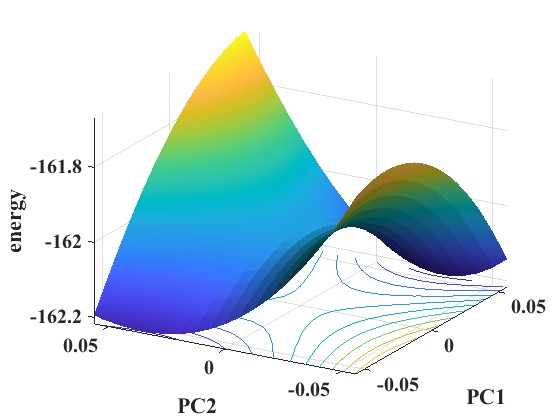}
    \caption{PC1, PC2}
    \end{subfigure}
   \begin{subfigure}[b]{0.45\textwidth}
    \centering
    \includegraphics[width=\textwidth]{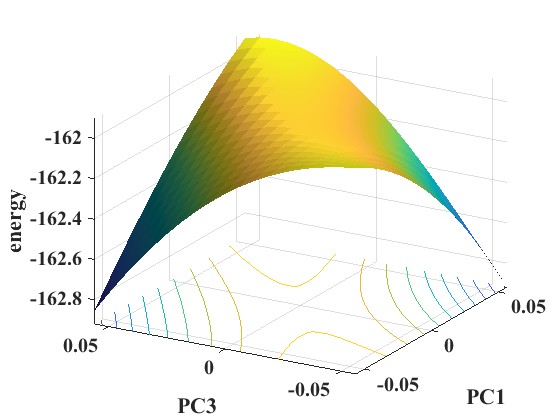}
    \caption{PC1, PC3}
    \end{subfigure}

    \begin{subfigure}[b]{0.45\textwidth}
    \centering
    \includegraphics[width=\textwidth]{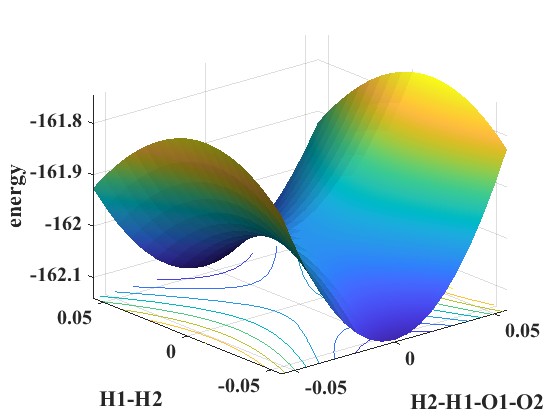}
    \caption{IC6(H2-H1-O1-O2), IC3(H1-H2)}
    \end{subfigure}
   \begin{subfigure}[b]{0.45\textwidth}
    \centering
    \includegraphics[width=\textwidth]{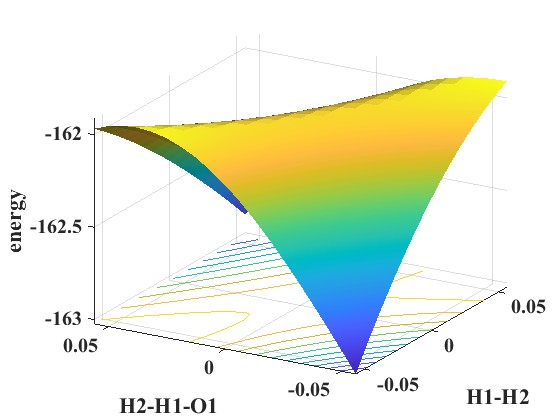}
    \caption{IC3(H1-H2), IC5(H2-H1-O1)}
    \end{subfigure}
    \caption{The potential energy surfaces of reaction 11 in pairs of PCs and pairs of ICs.}
     \label{fig:pes11}
\end{figure}

For reaction 11, Table~\ref{tab:corr}(b) shows that PC1 and PC2 are highly correlated with DC2 and DC3, respectively, indicating that they are good CVs. This observation is consistent with the PES shown in Figure~\ref{fig:pes11}(a) where a saddle point can be clearly seen at the transition state. Furthermore, it appears that the main reaction coordinate is mostly aligned with PC1, although it is not strictly parallel to PC1, and PC2 also contributes to the reaction coordinate.  We see from Table~\ref{tab:corr}(b) that the correlation coefficients between PC3 and DC2, and between PC3 and DC3 are relatively smaller, indicating that PC3 may not be a good CV. This is also evident from Figure~\ref{fig:pes11}(b) which shows that no clear saddle point can be seen from the PES plotted in PC1 and PC3. Table~\ref{tab:corr} also shows that both IC3 (H1-H2 bond length) and IC6 (H2-H1-O1-O2 dihedral angle) are highly correlated with DC2, and IC5 is moderately correlated with both DC2 and DC3. We see in Figure~\ref{fig:pes11}(c) that a saddle point is present in the PES plotted in IC3 and IC6, and the main reaction coordinate appears to be well described by IC3 (H1-H2 bond length), which is also the largest component in PC1 in magnitude. No clear saddle point is observed in Figure~\ref{fig:pes11}(d) where the PES is plotted in IC3 and IC5. These plots are consistent with the observed correlation coefficients between different ICs and DCs. They indicate that ICs that are highly correlated with DCs are indeed good CVs.

\begin{figure}[htbp]
   \centering
   \begin{subfigure}[b]{0.45\textwidth}
    \centering
    \includegraphics[width=\textwidth]{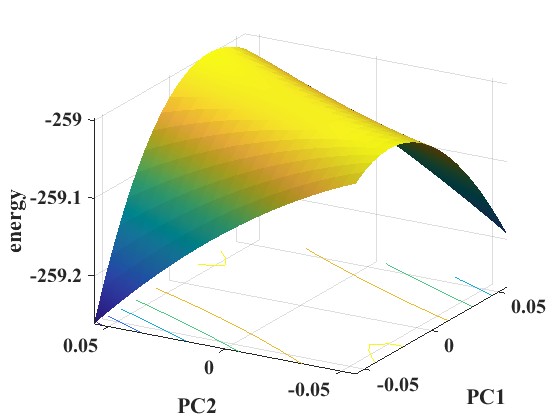}
    \caption{PC1, PC2}
    \end{subfigure}
   \begin{subfigure}[b]{0.45\textwidth}
    \centering
    \includegraphics[width=\textwidth]{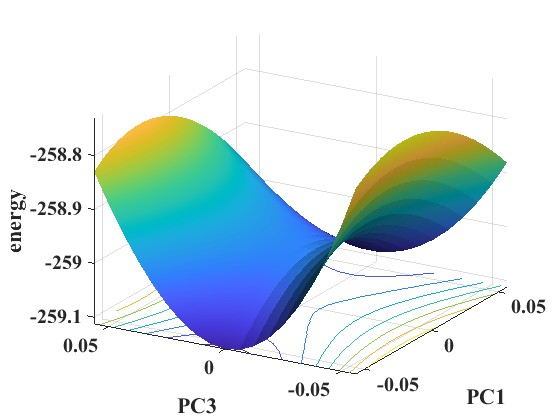}
    \caption{PC1, PC3}
    \end{subfigure}

    \begin{subfigure}[b]{0.45\textwidth}
    \centering
    \includegraphics[width=\textwidth]{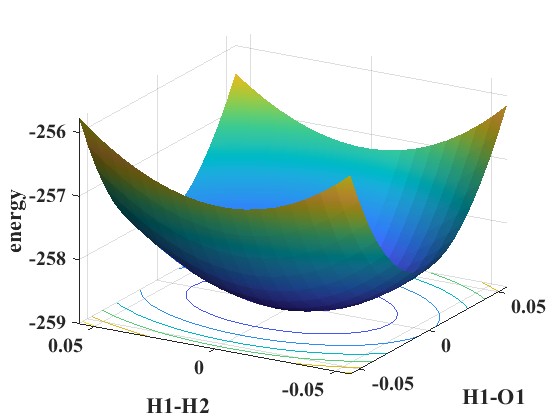}
    \caption{IC1(H1-O1), IC3(H1-H2)}
    \end{subfigure}
   \begin{subfigure}[b]{0.45\textwidth}
    \centering
    \includegraphics[width=\textwidth]{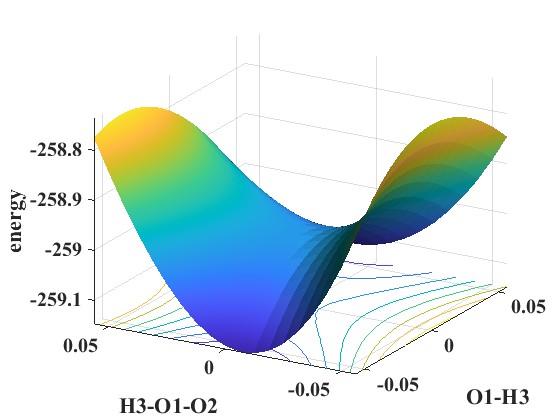}
    \caption{IC4(O1-H3), IC7(H3-O1-O2)}
    \end{subfigure}
    \caption{The potential energy surfaces of reaction 16 in pairs of PCs and pairs of ICs. 
    }
    \label{fig:pes16}
\end{figure}

Table~\ref{tab:corr}(c) shows that PC1 and PC3 are highly correlated with DC2 and DC3 for reaction 16, which is somewhat surprising. However, this may be explained by the observation that second and third singular values of the mean subtracted snapshot matrix are $\sigma_2 = 2.45$ and $\sigma_3 = 1.55$, which are very close.  Both are an order of magnitude smaller than $\sigma_1 = 24.28$ and 2-3 times larger than $\sigma_4 = 0.83$. Figure~\ref{fig:pes16}(b) shows that PC1 and PC3 are indeed better CVs to describe the reaction mechanism. In fact, the main reaction coordinate seems to be along the PC1 direction. Such information cannot be obtained from the PES plotted in PC1 and PC2 shown in Figure~\ref{fig:pes16}(a) where no saddle point can be found. Table~\ref{tab:corr}(c) also shows that IC4 (O1-H3 bond length), IC7(H3-O1-O2 torsion angle) and IC9 (H3-O1-O2-H2 dihedral angle) have relatively high correlations with either DC2 or DC3 indicating that they may be good CVs. Figure~\ref{fig:pes16}(d) confirms that the main reaction coordinate appears to be in the direction of IC4 (O1-H3 bond length) with some contribution from IC7 (H3-O1-O2 angle). A saddle point can clearly be identified at the transition state.  On the contrary, no reaction mechanism can be inferred from the PES plotted in IC1 and IC3 in Figure~\ref{fig:pes16}(c). Both IC1 and IC3 have low correlations with DC2 and DC3.

\begin{figure}[htbp]
   \centering
   \begin{subfigure}[b]{0.45\textwidth}
    \centering
    \includegraphics[width=\textwidth]{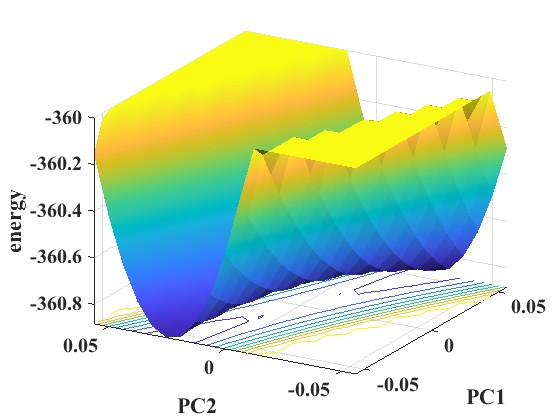}
    \caption{PC1, PC2}
    \end{subfigure}
   \begin{subfigure}[b]{0.45\textwidth}
    \centering
    \includegraphics[width=\textwidth]{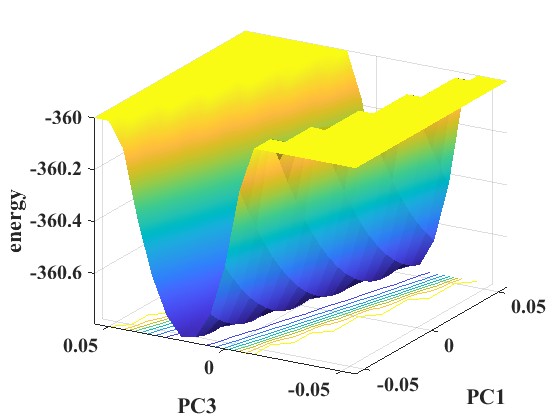}
    \caption{PC1, PC3}
    \end{subfigure}

    \begin{subfigure}[b]{0.45\textwidth}
    \centering
    \includegraphics[width=\textwidth]{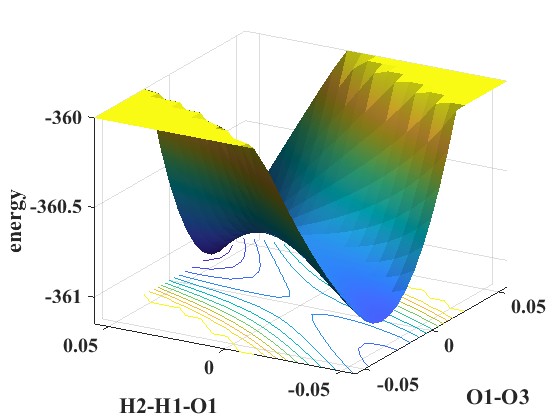}
    \caption{IC4(O1-O3), IC7(H2-H1-O1)}
    \end{subfigure}
   \begin{subfigure}[b]{0.45\textwidth}
    \centering
    \includegraphics[width=\textwidth]{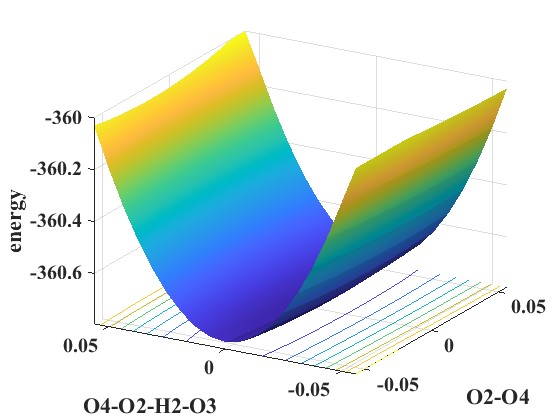}
    \caption{IC12(O4-O2-H2-O3), IC5(O2-O4)}
    \end{subfigure}
    \caption{The potential energy surfaces of reaction 14 in pairs PCs and pairs of ICs. 
    }
     \label{fig:pes14}
\end{figure}

Finally, Table~\ref{tab:corr}(d) shows that PC1 and PC2 are highly correlated with DC2 and DC3  respectively for reaction 14, indicating that they may be good CVs. This prediction is confirmed in Figure~\ref{fig:pes14}(a) in which the PES is plotted in PC1 and PC2. A saddle point can be observed at the full-dimensional transition state. However, the main reaction coordinate does not appear to be aligned with either PC1 or PC2 but a combination of them.  For this reaction, the first 4 singular values of the mean subtracted snapshot matrix are $\sigma_1 = 9.41$, $\sigma_2 = 5.59$, $\sigma_3 = 3.99$ and $\sigma_4=2.26$. Because $\sigma_1$ is not significantly larger than $\sigma_2$, both PC1 and PC2 (and possibly PC3 also) are important in describing the main reaction coordinate.
We see from Figure~\ref{fig:pes14}(b) that no saddle point can be clearly observed in the PES plotted in PC1 and PC3, which indicates that PC2 is a better CV than PC3, an oservation that is consistent with the conclusion drawn from the correlation coefficients reported in Table~\ref{tab:corr}.  

Reaction 14 has many more degrees of freedom (12) compared to other reactions considered earlier.  In Figure~\ref{fig:pes14}(c), we plot the PES in IC4 (O1-O3 bond length), which is highly correlated with DC3, and IC7(H2-H1-O1 angle) which is highly correlated with DC2, as we can see from Table~\ref{tab:corr}. A saddle point can clearly be seen at the transition state indicating that both IC4 and IC7 are good CVs as predicted by their high correlations with DC3 and DC2 respectively. The main reaction coordinate does not seem to be aligned with either one of them, but a combination of the two.  On the contrary, no saddle point can be observed in Figure~\ref{fig:pes14}(d) where the PES is plotted in IC12 (O4-O2-H2-O3 dihedral angle) and IC5 (O2-O4 bond length). Neither one of these ICs is highly correlated with DC2 or DC3 as we can see from Table~\ref{tab:corr}(d).

\subsection{Committor Analysis}\label{sec:commanalysis}
\begin{figure}[htbp]
   \centering
     \begin{subfigure}[b]{0.45\textwidth}
    \centering
    \includegraphics[width=\textwidth]{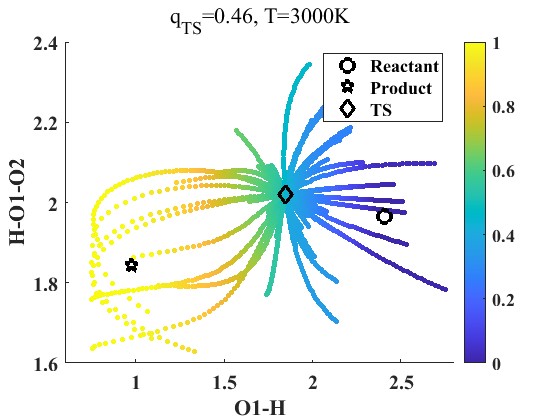}
    \caption{Committor probabilities}
    \end{subfigure}
   \begin{subfigure}[b]{0.45\textwidth}
    \centering
    \includegraphics[width=\textwidth]{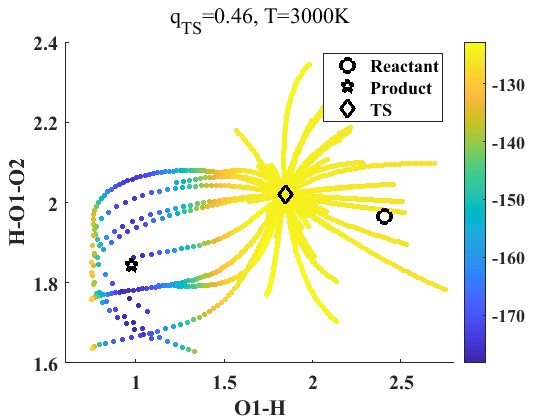}
    \caption{Free energies}
    \end{subfigure}
    \caption{Committor analysis for reaction 9. (a) Committor probabilities and (b) free energies at 3000K are plotted in the space of (IC2, IC3) listed in Table~\ref{tab:corr}. We used Eq. \eqref{eq:kernelmat1} to construct the diffusion map with $\alpha=\frac{1}{2}$ and computed the committor probabilities.}  
    \label{fig:committors09}
\end{figure}

\noindent
We use the AIMD dataset in \cite{hcombustdata22} to build global diffusion maps and compute committor functions as described in \eqref{eq:committor}.  In our experiments, we set $r=0.002$ in \eqref{knearest} to determine local scale parameters and construct a global diffusion map from snapshots at a given temperature. We verified that values of the computed committor functions at transition states are close to 0.5. Figure~\ref{fig:committors09}(a) shows the computed committor function for reaction 9 evaluated at configurations along AIMD trajectories generated at $T=3000K$.  
The committor function is plotted in the plane formed by two internal coordinates IC2 (O1-H distance) and IC3 (H-O1-O2 angle) that are highly correlated with the first diffusion coordinate as reported in  Table~\ref{tab:corr}. The variation of the IC1 (O1-O2 distance) among all the snapshots is less than 0.16\AA, therefore the 2D view of the committor function in Figure~\ref{fig:committors09}(a) fully captures the main features of committor function. We can clearly see that the values of the committor function are closer to 1 to the left of the transition state, while they are closer to 0 to the right of the transition state. 

We also plot the projected locations of the reactant and product in this two dimensional space and observe that they are in reactant and product regions that are clearly separated by the transition state. From this figure, we can clearly see that O1-H distance is globally the main reaction coordinate, which is consistent with the observation we made in Figure~\ref{fig:pes09}(d) in a local region around the transition state. We also see that the information provided in the committor function is consistent with the free energy surface plot shown in Figure~\ref{fig:committors09}(b) in which the configurations to the left of the transition state have much lower energies, whereas the configurations to the right of the transition state have slightly lower energies compared to the free energy at the transition state.


\begin{figure}[htbp]
   \centering
   \begin{subfigure}[b]{0.45\textwidth}
    \centering
    \includegraphics[width=\textwidth]{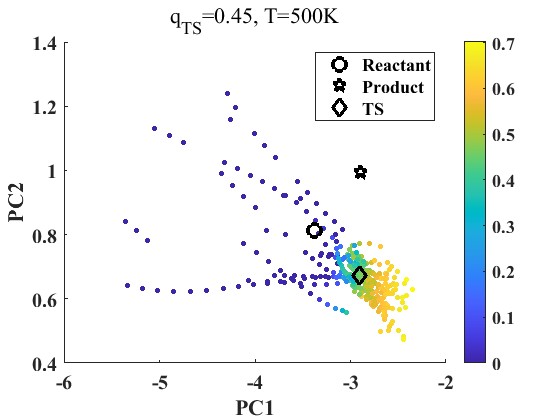}
    \caption{Committor probabilities}
    \end{subfigure}
   \begin{subfigure}[b]{0.45\textwidth}
    \centering
    \includegraphics[width=\textwidth]{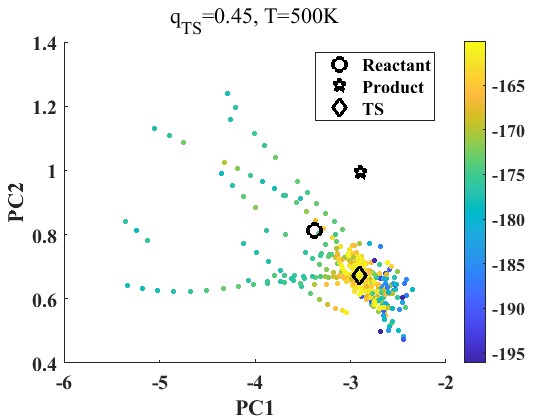}
    \caption{Free energies}
    \end{subfigure}
    \caption{Committor analysis for reaction 11. (a) Committor probabilities and (b) free energies at 500K are plotted in the space of (PC1, PC2). We used Eq. \eqref{eq:kernelmat1} to construct the diffusion map with $\alpha=\frac{1}{2}$ and computed the committor probabilities.}
     \label{fig:committors11}
\end{figure}

Figure~\ref{fig:committors11}(a) shows the committor functions in the PC1-PC2 plane at selected configurations along the AIMD trajectories of reaction 11 generated at $T=500K$. The configurations are selected by restricting the coefficients of the other 4 PCs (out of a total of 6) to be close to those obtained at the transition state. We again observe that the committor function has similar behavior near the transition state as before, and that in this case PC1 is the main reaction coordinate as is consistent with the potential energy surface plot shown in Figure~\ref{fig:pes11}(b). However, we also observe that away from the transition state, the main reaction coordinate is not completely determined by PC1 as observed when we plot the projected locations of the reactant and product in the PC1-PC2 plane, in which the reactant and product PC coefficients are not close to those at the transition state. Therefore, the plotted locations of the reactant and product in the PC1-PC2 plane do not completely describe their proximity to the transition state in this figure. Because the committor function is only evaluated at configurations along the AIMD trajectory, we do not have the committor function value at configurations in which the PC1 and PC2 coefficients are fixed at those associated with the reactant or product while the coefficients of all other PCs' are fixed at those associated with the transition state.

\begin{figure}[htbp]
    \centering
     \begin{subfigure}[b]{0.45\textwidth}
    \centering
    \includegraphics[width=\textwidth]{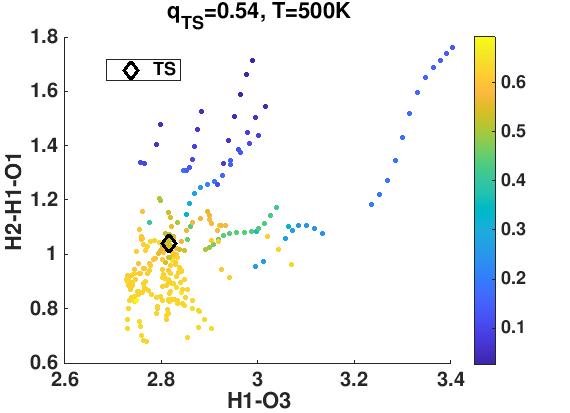}
    \caption{Committor probabilities}
    \end{subfigure}
   \begin{subfigure}[b]{0.45\textwidth}
    \centering
    \includegraphics[width=\textwidth]{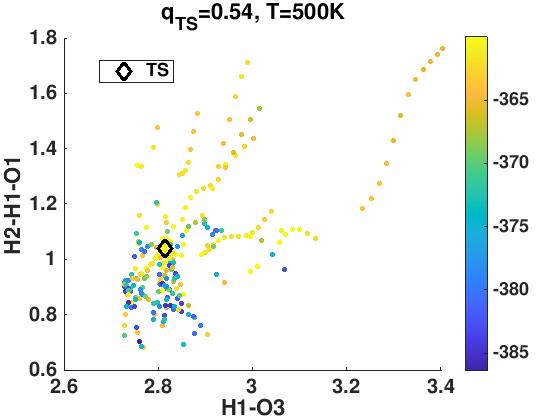}
    \caption{Free energies}
    \end{subfigure}
    \caption{Committor analysis for reaction 14. (a) Committor probabilities and (b) free energies are shown in the space of (IC4, IC7) listed in Table~\ref{tab:corr}. Snapshots near the TS are thresholded with respect to the ICs so that the 2-dimensional projection of snapshots is performed approximately. We used Eq. \eqref{eq:kernelmat2} to construct the diffusion map and computed the committor probabilities. For reaction 14, the committor value at the TS is close to 0.5 only at 500K.}
    \label{fig:committors14}
\end{figure}
      
In the case of reaction channel 14, the selected configurations with ICs close to those at the transition state except IC4 and IC7 have committor function values close to zero to the upper right of the transition state and close to one to the lower left of the transition state in the IC4-IC7 plane as shown in Figure~\ref{fig:committors14} (a). This is consistent with the potential energy surface around the transition state plotted in Figure~\ref{fig:pes14} (b). Near the transition state, the main reaction coordinate appears to be mostly determined by IC7 with a small contribution from IC4. However, away from the transition state, the main reaction path is determined by a different linear combination of IC4 and IC7 depending on the location of the path.  

\begin{figure}[htbp]
   \centering
   \begin{subfigure}[b]{0.45\textwidth}
    \centering
    \includegraphics[width=\textwidth]{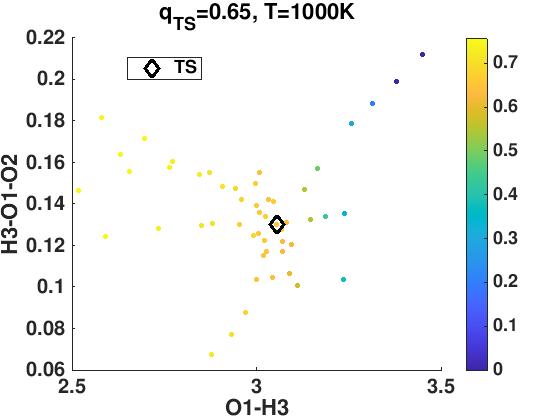}
    \caption{Committor probabilities}
    \end{subfigure}
   \begin{subfigure}[b]{0.45\textwidth}
    \centering
    \includegraphics[width=\textwidth]{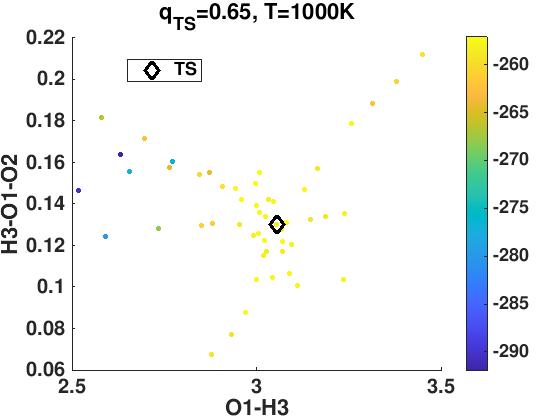}
    \caption{Free energies}
    \end{subfigure}

    \begin{subfigure}[b]{0.45\textwidth}
    \centering
    \includegraphics[width=\textwidth]{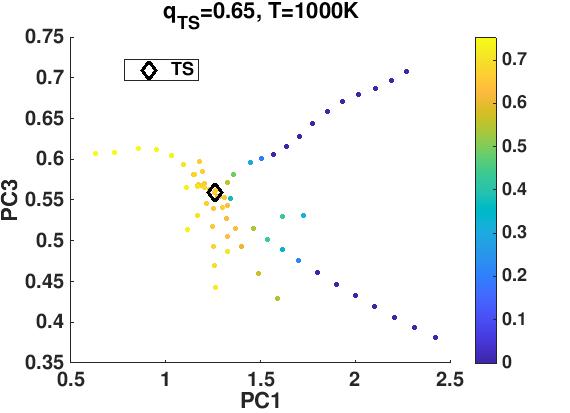}
    \caption{Committor probabilities}
    \end{subfigure}
   \begin{subfigure}[b]{0.45\textwidth}
    \centering
    \includegraphics[width=\textwidth]{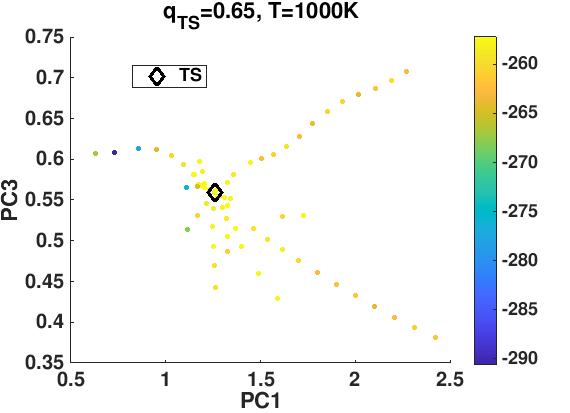}
        \caption{Free energies}
    \end{subfigure}

    \caption{Committor analysis for reaction 16. (a) Committor probabilities and (b) free energies at 1000K are projected onto in the spaces of (IC4, IC7) listed in Table~\ref{tab:corr} and (PC1, PC3). We used Eq. \eqref{eq:kernelmat2} to construct the diffusion map and computed the committor probabilities.}
     \label{fig:committors16}
\end{figure}

Figure~\ref{fig:committors16} shows the committor function and free energies of selected molecular configurations along the AIMD trajectories associated with reaction 16 at $T = 1000K$. The configurations are selected to either have their ICs close to those associated with the transition state except IC4 and IC7 or have their PC coefficients close to those associated with the transition state except the coefficients of PC1 and PC3. The committor function and free energy are plotted in the plane of IC4 and IC7 and in the plane of PC1 and PC3 which are determined to be good CVs based on their correlation with the first two diffusion coordinates shown in Table~\ref{tab:corr}. This is consistent with the corresponding potential energy plot near the transition state shown in Figure~\ref{fig:pes16} (d) although the number of configurations shown in these plots is limited.

\section{Conclusion}
Inspired by the recent work presented in~\cite{tlldiffmap20}, we considered the use of diffusion maps to analyze several reaction channels involved in a hydrogen combustion system. In particular, we use local diffusion maps constructed from configurations sampled along AIMD trajectories within a metastable region to identify and assess CVs obtained from internal coordinates and principal component analysis. 

We found that the correlations between the first few principal components and the first two diffusion coordinates tend to be high. The two PCs that have the highest correlations with respect to the first two diffusion coordinates were found to be good CVs for characterizing the reaction path near the transition state, and validated by the presence of a saddle point on the potential energy surface. We also observed that, while several ICs can be highly correlated with the first two diffusion coordinates, they are not uniformly all good CVs. Nonetheless, the use of diffusion maps allows us to narrow down the IC choices and provides some alternatives that we otherwise would not have considered.

By using AIMD trajectories initiated from the transition states, we also constructed global diffusion maps that can be used to compute approximations to the committor functions associated with different reaction channels.  When we examined the committor function in the plane of two CVs identified by a diffusion map, we found that for all reaction channels the value of the committor function at the transition state is close to 0.5. For reaction channels that contain only a few degrees of freedom, the reactant and product are clearly separated by the selected CVs, and the main reaction path characterized by the committor function agrees well with that identified from the free energy surface in the selected CVs. However, for reaction channels that involve more degrees of freedom, the global reaction coordinate appears to be more complicated, and the reactant and product may not be easily separated on a two-dimensional slice of committor function defined in a high dimensional space. Even so, insights into the CVs provided by the diffusion maps have proved useful enough such that we will report the transition state free energies found for all hydrogen combustion reaction channels in a related publication.

\section*{Acknowledgement}

This work was supported by the U.S. Department of Energy, Office of Science, Office of Advanced Scientific Computing, and Office of Basic Energy Sciences, via the Scientific Discovery through Advanced Computing (SciDAC) program. This work used computational resources provided
by the National Energy Research Scientific Computing Center (NERSC), a U.S. Department of Energy Office of Science User Facility operated under Contract DE-AC02-05CH11231. 

\bibliography{achemso-demo}

\end{document}